\documentclass[12pt,preprint]{aastex}

\shorttitle{A Flare Model for Sgr A*}

\begin{document}


\title{A Testable Stochastic Acceleration Model \\ for Flares in Sagittarius A*}


\author{Siming Liu,\altaffilmark{1} Vah\'e Petrosian,\altaffilmark{2} 
Fulvio Melia,\altaffilmark{3} and Christopher L. Fryer\altaffilmark{1, 4}
}

\altaffiltext{1}{Los Alamos National Laboratory, Los Alamos, NM 87545; liusm@lanl.edu}
\altaffiltext{2}{Center for Space Science and Astrophysics, Department of Physics and Applied Physics, Stanford 
University, Stanford, CA 94305; vahe@astronomy.stanford.edu}
\altaffiltext{3}{Physics Department and Steward Observatory, The University of Arizona, 
Tucson, AZ 85721; melia@physics.arizona.edu; Sir Thomas Lyle Fellow and Miegunyah Fellow.}
\altaffiltext{4}{Physics Department, The University of Arizona, Tucson, AZ 85721; fryer@lanl.gov}


\begin{abstract}

The near-IR and X-ray flares in Sagittarius A* are believed to be produced by relativistic electrons 
via synchrotron and synchrotron self-Comptonization, respectively. These electrons are likely 
energized by turbulent plasma waves through second order Fermi acceleration that, in combination 
with the radiative cooling processes, produces a relativistic Maxwellian distribution in the steady 
state. This model has four principal parameters, namely the magnetic field $B$, the electron density $n$ and 
temperature $\gamma_c\,m_e c^2$, and the size of the flare region $R$. In the context of 
stochastic acceleration, the quantities $R\,n^{1/2} B$ and $\gamma_c R n$ should remain 
nearly constant in time. Therefore, simultaneous spectroscopic observations in the NIR and X-ray bands 
can readily test the model, which, if proven to be valid, may be used to determine the evolution of 
the plasma properties during an eruptive event with spectroscopic observations in either band or 
simultaneous flux density measurements in both bands. The formulae can be applied to other 
isolated or confined systems, where electrons are accelerated to relativistic energies by plasma wave turbulence 
and produce most of the emission via synchrotron processes.

\end{abstract}





\section{Introduction}

Sagittarius A*, the compact radio source at the Galactic Center, is powered by 
accretion onto an $M\sim 3-4\times 10^6\;M_\odot$ supermassive black hole 
(Sch\"{o}del et al. 2002; Ghez et al. 2004; Melia 2006). There is compelling 
evidence that the near-IR (NIR) flares in Sagittarius A* are produced by relativistic 
electrons via synchrotron emission within a few Schwarzschild radii of the 
black hole (Genzel et al. 2003; the Schwarzschild radius $r_S =10^{12}(M/3.4\times 
10^6\;M_\odot$) cm). Inverse Compton 
scattering of the NIR photons 
by the same electrons produces X- and $\gamma$-ray emission, which may account 
for the observed X-ray flares (Baganoff et al. 2001, 2003; Goldwurm et al. 2003; 
Porquet et al. 2003; B\'{e}langer et al. 2006). Simultaneous multi-wavelength 
observations during a handful of flares have generally supported this hypothesis 
(Eckart et al.  2004, 2005; B\'{e}langer et al. 2005; Yusef-Zadeh et al. 2006). 
However, our knowledge of the flare energization and electron acceleration 
mechanisms is still very rudimentary, though some details are starting to emerge 
(Liu et al. 2004, 2006; Tagger \& Melia 2006; Bromley, Melia, \& Liu 2001; 
Broderick \& Loeb 2005).

The first NIR spectroscopic observations at a frequency $\nu \simeq 1.4\times 
10^{14}$ Hz revealed very soft power spectra, $\nu F_{\nu}\propto \nu^{\alpha}$, 
with index $-4<\alpha<-2$ (Eisenhauer et al. 2005), indicating a sharp cutoff in 
the electron distribution. Such a cutoff is a natural consequence of stochastic 
acceleration (SA) by plasma waves, in which low-energy electrons are accelerated 
up to a relativistic energy, where radiative cooling becomes important (Schlickeiser 
1984; Park \& Petrosian 1995).  In Sagittarius A*, the X-ray flare luminosity is always lower 
than that of the NIR events, so synchrotron cooling dominates. More recent NIR 
observations also indicate that $F_\nu$ may be correlated with $\alpha$ (Ghez et 
al. 2005; Gillessen et al. 2006; Hornstein et al. 2006).

Flares in Sagittarius A* are likely driven by an accretion instability (e.g., Tagger \& Melia 
2006), which triggers the transfer of gravitational energy of the plasma into turbulence. In earlier 
work, we demonstrated that if thermal synchrotron and synchrotron self-Comptonization are the 
dominant emission mechanisms during Sagittarius A*'s flares, simultaneous X-ray and NIR spectroscopic 
observations (SSO) can measure the size of the flare region $R$, the electron density $n$ and 
``temperature'' $\gamma_c m_e c^2$, and the magnetic field $B$ (Liu, Melia, \& Petrosian 2006). 
Interestingly, most of the flares are found to be local events with the source size $R<r_S$, which can 
also explain the quasi-period modulations observed in some flares (see also Liu, 
Petrosian, \& Melia 2004). This justifies the neglect of the source structure as a 
zero order approximation and suggests that the flares involve highly dynamical 
activity and the electron acceleration process has to be addressed to uncover the 
underlying physical processes.

In this paper, we consider steady state electron populations produced via SA by turbulent plasma 
waves which, in combination with the synchrotron cooling, results in a relativistic Maxwellian distribution. 
The electron temperature is set by the energy diffusion and synchrotron cooling rates. 
The latter depends only on $B$, while the former is given by a combination of $B$, $n$, 
and $R$, which together imply that the product $\gamma_c n R$ should not vary 
significantly as the flare evolves, nor from one flare to the next. Moreover, since 
electrons gain energy from the plasma waves, overall energy conservation requires that the 
luminosity be equal to the energy transfer rate from turbulence, which also depends only 
on $B$, $n$, and $R$. This requires $B n^{1/2} R$ to be constant. These two system 
constraints effectively reduce the model parameters to just two, so that SSO can readily 
test the model.

Previous work on the flares is summarized in \S\ \ref{bgd}, where we discuss difference between these 
models and assess the merit of previous proposed scenarios on the definiteness of their predictions and/or
whether the model parameters can be well constrained by observations. The electron acceleration is described 
in \S\ \ref{acc}. And in \S\ \ref{apl} the model is applied to an NIR flare with a very hard 
spectrum. The model predictions and its limitations are discussed in \S\ \ref{dis}.

\section{Previous Work}
\label{bgd}

Several models have been proposed since the {\it Chandra} discovery of a bright X-ray flare from 
Sagittarius A* (Baganoff et al. 2001; Markoff et al. 2001; Liu \& Melia 2002; Yuan et al. 2004; 
Nayakshin, Cuadra, \& Sunyaev 2004). The facts that the flare variation time scale is comparable to 
the orbital period at the last stable orbit and there is a good correlation between the X-ray and the 
strongly polarized NIR emission suggests that the flares are produced very close to the black hole. 
The nature of the correlation between an X-ray flare and a delayed radio outburst, on the other hand, indicates 
that the radio emission may be produced at large radii (Zhao et al. 2004). In the jet model studied by Markoff 
et al. (2001), it was assumed that the electrons and protons are Maxwellian and have the same 
temperature, the plasma near the base of the jet, which is within a few Schwarzschild radii of the 
black hole, are heated suddenly to a higher temperature during flares via shocks or magnetic 
field reconnection. By changing the physical conditions at the base of the jet, they were able to 
reproduce the observed X-ray flux. The distinct spectral and polarization characteristics of 
sub-millimeter emission, on the other hand, suggests that this radiation is produced by a small hot 
magnetized accretion torus near the black hole (Melia et al. 2000, 2001). The flat spectrum of the 
X-ray flares then motivated us to consider the production of X-rays via bremsstrahlung during an 
accretion instability (Liu \& Melia 2002). To reduce the model parameters, the electrons and protons 
were assumed in thermal equilibrium in this calculation. Yuan et al. (2004) later considered a more 
complicated accretion model, where the electrons and protons have different temperatures and there is 
a non-thermal component in the electron distribution prescribed by the choice of particle acceleration 
processes. With the temperature for the thermal component, the spectral index, high energy cutoff, 
and normalization factor characterizing the non-thermal component as free parameters, it can accommodate a 
variety of flare spectra. 

All these models are still consistent with the limited NIR and X-ray observations, due in part to the 
large number of model parameters and the relative paucity of data. It is clear that one must address
the electron acceleration mechanism in order to (possibly) distinguish between these various models. 
Given the physical conditions in the emission region---those of a hot, magnetized plasma---it is natural 
to suppose that the electrons are being accelerated by plasma waves (e.g., Dermer, Miller, \& Li 1996), 
whose turbulent energy is fed from an accretion-induced instability. However, given the uncertainties in the 
generation of turbulence, and in the cascade and damping processes, several parameters were introduced to 
describe the turbulence spectrum (Liu et al. 2004). Not surprisingly, as was the
case for the previous models, the available observations could not provide rigorous constraints on the
parameters. Nevertheless, this first attempt at building a physical model suggested that some of the 
flares are actually produced within a region smaller than the size of the black hole, indicating that 
they are likely events localized in certain portions of the disk.

The discovery of NIR flares with very soft spectra implies a sharp cutoff in the relativistic electron 
distribution producing the polarized NIR emission via synchrotron. For typical conditions near the black 
hole inferred from the modeling of Sagittarius A*'s quiescent emission (Melia et al. 2001; Yuan et al. 
2003), the synchrotron cooling time of electrons producing NIR emission is also comparable to the flare 
variation time scale and may therefore be the cause of the cutoff. Subsequent work with this model showed that 
the electron acceleration may be insensitive to the details of the turbulence spectrum and that such a 
cutoff is a natural consequence of SA of electrons by turbulence (Liu et al. 2006). Moreover, if 
the injection of particles into the acceleration region is slow, the electron spectrum in steady
state may be approximated as a relativistic Maxwellian distribution. In this case, there are only four model 
parameters, namely the source size $R$, the electron temperature $\gamma_c m_ec^2$ and density $n$, and the 
magnetic field $B$. This model has therefore now matured to the point where its parameters may be readily 
determined by SSO. Of course, this does not yet mean that SA is solely responsible for accelerating 
the electrons. Other mechanisms may also produce similar distributions.

\section{SA of Electrons by Plasma Waves}
\label{acc}

Tagger and Melia (2006) first showed that the Rossby wave instability in a small accretion torus can 
reproduce the characteristics of the flare light curves. In such a picture the gravitational energy of 
protons and ions is released by this macroscopic instability, which generates the turbulence on a scale 
comparable to $R$. The turbulence then cascades toward small scales and accelerates electrons in the 
process. It is clear that the whole plasma in the turbulent region can be energized by the gravitational 
energy release. We therefore do not need a continuous injection of electrons at low energies. Moreover, 
since the plasma is gravitationally trapped by the black hole, the accelerated electrons will not escape from 
the acceleration region as long as its energy is lower than the gravitational banding energy of the protons, 
which is comparable to the protons rest mass energy. The evolution of an electron distribution $N(\gamma,t)$ 
under the influence of a turbulent magnetic field is given therefore by (Blandford \& Eichler 
1987)
\begin{equation} 
   {\partial N \over \partial t}
=  {\partial \over \partial \gamma}\left[{\gamma^4\over \tau_{\rm ac}}{\partial \gamma^{-2}N \over \partial 
\gamma} + {\gamma^2\over \tau_0} N\right]\,,
\label{kinetic}
\end{equation}
where $\gamma=100 \gamma_2$ is the Lorentz factor\footnote{Here we have assumed that the direct acceleration 
rate $<{\Delta \gamma}/\Delta t>/\gamma$ is equal to the energy diffusion rate $ 
<\Delta\gamma\Delta\gamma/\Delta t>^\prime/\gamma$ as is usually true in SA by plasma waves, where a 
prime indicates a derivative with respect to $\gamma$. In a more general case, where the direct 
acceleration rate $<\Delta\gamma/\Delta t>/\gamma = p <\Delta\gamma\Delta\gamma/\Delta 
t>^\prime/\gamma$ the first term on the right hand side of equation (\ref{kinetic}) should be 
$[\gamma^{4p}\tau_{\rm ac}^{-1}(\gamma^{2-4p}N)^\prime]^\prime$.}
and the synchrotron cooling
and acceleration times are given, respectively, by,
\begin{eqnarray}
\tau_{\rm syn}(\gamma) &=&\tau_0/\gamma
\equiv {9m_e^3c^5/4 e^4B^2\gamma}
=  21.56\ B_1^{-2} \gamma_2^{-1}\ {\rm hrs}\,, \\ 
\tau_{\rm ac} &\equiv& 2\gamma^2/<\Delta\gamma\Delta\gamma/\Delta
t> = C_1{3R c/v_{\rm A}^{2}} 
= 52.49\ C_1 R_{12} n_7 
B_1^{-2}\ {\rm hrs}\,,
\label{acct}
\end{eqnarray}
where $v_{\rm A} = B/(4\pi n m_p)^{1/2} =6.901\times 10^8B_1 n_7^{-1/2}$cm s$^{-1}$ is the Alfv\'{e}n 
velocity, $B_1 = B/10$ Gauss, $n_7 = n/10^7$ cm$^{-3}$, $R_{12} = 
R/10^{12}$ cm, and $C_1$ is a dimensionless quantity that depends on the microscopic details 
of the wave-particle interaction, and is of order 1 if the scattering mean-free-path is 
comparable to $R$, which may be smaller than the Schwarzschild radius ($\sim 10^{12}$ cm) 
of the black hole. The other constants have their usual meaning. Note that in equation (\ref{acct}) 
the velocity of the scatterers in the original 
SA proposed by Fermi (1949) has been replaced by the Alfv\'{e}n velocity $v_{\rm A}$. This may not be always 
true especially when the sound velocity $v_s>v_{\rm A}$. The exact dependence of $\tau_{\rm ac}$ on $v_s$ and 
$v_{\rm A}$ is still under investigation. Because $v_s$ also depends on the proton temperature, which is 
an input in the model, here we focus on the simplest scenario, where the effects of sound waves are 
ignored. And as we shall see below, particle-particle collisions can be ignored, and the electrons 
are mainly scattered by the turbulence. Since the inferred particle pressure is much higher than the magnetic 
field pressure, $R$ is the only length scale in the system and should be comparable to the particle 
mean-free-path.
If the flares are triggered by instabilities that are similar from event to event, $C_1$ should be 
nearly constant. In steady state this yields\footnote{For the more general case mentioned above $N(\gamma) 
\propto \gamma^{4p-2}\exp(-\gamma/\gamma_c)$.}
\begin{equation}
N(\gamma)= (n\gamma^2/ 2\gamma_c^3)\exp(-\gamma/\gamma_c)\,,\ \ \ \ \ \, {\rm with} \ \ \ \ \ \gamma_c 
= \tau_0/\tau_{\rm ac}=41.08\ C_1^{-1}R_{12}^{-1}n_7^{-1}\,.
\label{gamc}
\end{equation}
The normalization of $N$ is accurate to order $\gamma_c^{-2}$.
Therefore $\gamma_c R\, n$ only 
depends on the instabilities and should not change significantly with time.

\section{Relativistic Maxwellian Synchrotron Emission and Self-Comptonization}
\label{apl}

The thermal synchrotron flux density and emission coefficient (i.e. emissivity divided by the solid angle) are 
given, respectively, by (Petrosian 1981; Mahadevan, Narayan, \& Yi 1996)
\begin{equation}
F_{\nu}=(4\pi R^3/ 3 D^2){\cal E}_\nu\,,\ \ \ \
{\cal E}_\nu=(\sqrt{3} e^3/ 8\pi m_e c^2) B\, n\, x_M\, I(x_M)\,,
\end{equation}
where 
\begin{eqnarray}
I(x_M)&=& 4.0505x_M^{-1/6}(1+0.40x_M^{-1/4} + 0.5316x_M^{-1/2})\exp(-1.8899\,x_M^{1/3})\,,
\label{Im}
\\ 
x_M &=& {\nu/\nu_c} 
\equiv {4\pi m_e c\nu/3 e B \gamma_c^2} = 1412\ 
C_1^2\ \nu_{14} R_{12}^2\ n_7^2\ B_1^{-1}\,,
\label{xm}
\end{eqnarray}
$\nu =\nu_{14} 10^{14}$ Hz, and the distance to the Galactic Center $D = 
D_8 8$ kpc. The spectral index in a given narrow frequency 
range is therefore 
\begin{equation}
\alpha \equiv {{\rm d}\ln(\nu F_\nu)\over {\rm d}\ln\nu} = 1.833 - 0.6300 
x_M^{1/3}-{0.1000x_M^{1/4}+0.2658\over 
x_M^{1/2}+0.4000x_M^{1/4}+0.5316}\,,
\label{alpha}
\end{equation}
and the flux density may be written 
\begin{equation}
F_\nu = {e^3\over 2\sqrt{3} m_e c^2} {B n R^3 \over D^2} x_M I(x_M) = 639.7\ R_{12}^3\  n_7\ B_1\ 
D_8^{-2} x_M I(x_M)\ {\rm mJy}\,.
\label{flux}
\end{equation}
The thick lines in Figure \ref{fig1.ps} (left) exhibit this spectral index 
(dotted), and the normalized spectrum $x^2_M I(x_M)$ (solid), as functions of the 
normalized frequency $x_M$.

We can estimate the energy transfer rate in a fully developed turbulence from dimensional analysis: 
\begin{equation}
{\cal L}_{\rm Turb} = 
C_2 v_{\rm A} B^2 R^2 = 6.901 \times 10^{34}
C_2\ R_{12}^2\ n_7^{-1/2}  B_1^3\ {\rm ergs\ s}^{-1}\,,
\end{equation}
where $C_2$ is another dimensionless quantity that only depends on the turbulence 
generation mechanism and therefore remains nearly a constant\footnote{Here ${\cal L}_{\rm Turb}$ could be a 
complicated function of $v_s$ and $v_{\rm A}$. We, however, focus on the simplest case.}. If there are no 
other significant energy loss processes, this energy transfer rate should be equal to the 
source luminosity (${\cal L}\simeq {\cal L}_{\rm syn}$) in the steady state, where 
\begin{equation}
{\cal L}_{\rm syn}= (64\pi e^4/ 9m_e^2 c^3)\ n\ R^3 B^2\gamma_{c}^2
= 8.949\times 10^{34} C_1^{-2} R_{12}\ n_7^{-1}B_1^2\ {\rm ergs\ s}^{-1}\,.
\end{equation}
We therefore infer that
\begin{equation}
R_{12}\ n_7^{1/2} B_1 = 1.297\ C_1^{-2} C_2^{-1}\,,
\label{c2}
\end{equation}
a condition that should not change significantly with time either. We note that the electron 
advection occurs on the accretion time, which can be more than 10 times longer than the dynamical time. This 
term is negligible during flares (see \S\ \ref{dis}). And even though there are other energy loss processes, 
such as the diffusive escape of accelerated particles and/or the propagation of waves away from the flare 
region (Petrosian \& Liu 2004), we expect that an approximately fixed fraction of the released energy goes 
into radiation. 
This constraint is still valid.

For given values of $C_1$ and $C_2$, spectroscopic observations in the NIR band then can be used to 
determine the properties of the flaring plasma. First from equation (\ref{alpha}) and the observed spectral 
index $\alpha$, one can obtain $x_M$, which in combine with equations (\ref{xm}) and (\ref{Im}) and the 
observed frequency gives $\nu_c$ and $I(x_M)$. Then one can use the measured flux density and equations 
(\ref{xm}), (\ref{flux}), and (\ref{c2}) to obtain $R$, $n$, and $B$, and $\gamma_c$ can be 
inferred with equation (\ref{gamc}). For the bright NIR flare observed by Ghez et al. (2005), $\alpha = 
0.5$, $\nu_{14}=1.429$, and $F_\nu = 7$ mJy, we have
$x_{M} = 7.314\,,\   
x_{M} I(x_{M}) = 0.7811\,,\ 
\nu_c = 1.954\times 10^{13} 
({x_{M}/7.314})^{-1}
({\nu_{14}/1.429}) \ {\rm Hz}
\,,$
\begin{eqnarray}
R_{12}&=& 0.08471\ 
C_1^{2} C_2^{6/7} D_8^{10/7}
\left({x_{M}\over7.314}\right)^{-1/7}
\left({x_{M} I(x_{M})\over0.7811}\right)^{-5/7}
\left({\nu_{14}\over 1.429}\right)^{1/7} 
\left({F_\nu\over 7\ {\rm mJy}}\right)^{5/7} 
\,,\\
n_7 &=& 2.266\ C_1^{-4} C_2^{-10/7} D_8^{-12/7}
\left({x_{M}\over7.314}\right)^{4/7}
\left({x_{M} I(x_{M})\over0.7811}\right)^{6/7}
\left({\nu_{14}\over 1.429}\right)^{-4/7} 
\left({F_\nu\over 7\ {\rm mJy}}\right)^{-6/7} 
\,,\\
B_1 &=& 10.17\ 
C_1^{-2} C_2^{-8/7} D_8^{-4/7}
\left({x_{M}\over7.314}\right)^{-1/7}
\left({x_{M} I(x_{M})\over0.7811}\right)^{2/7}
\left({\nu_{14}\over 1.429}\right)^{1/7} 
\left({F_\nu\over 7\ {\rm mJy}}\right)^{-2/7} \,,\\
\gamma_c&=& 214.0\
C_1 C_2^{4/7} D_8^{2/7}
\left({x_{M}\over7.314}\right)^{-3/7}
\left({x_{M} I(x_{M})\over0.7811}\right)^{-1/7}
\left({\nu_{14}\over 1.429}\right)^{3/7} 
\left({F_\nu\over 7\ {\rm mJy}}\right)^{1/7}
\,.
\end{eqnarray}
All these values are consistent with the results of previous studies (Liu et al. 2004, 2006). The 
electron-electron collision time, $\tau_c = \gamma m_e^2 c^3/8\pi n 
e^4 \ln\Lambda \simeq 1164\ \gamma_2 n_7^{-1}$ hrs ($\ln \Lambda=40$ in this case), is much longer 
than any other time scale. Particle-particle collisions, therefore, may be ignored. The electron pressure 
($\sim n \gamma_c m_e c^2$) is about 5 times higher than 
the magnetic field pressure. The total particle pressure could be much higher.

From Kirchhoff's law, the synchrotron absorption coefficient $$\kappa_\nu = {{\cal E}_\nu/2 \gamma_c m_e 
\nu^2}
= (\pi e\ n/ 3\sqrt{3}\gamma_c^5B)[I(x_M)/ x_M]\,,$$ we have the optical depth through 
the emission region 
\begin{eqnarray}
\tau_\nu &\equiv& \kappa_\nu R = 2.482\ C_1^5\ R_{12}^6\ n_7^6\ B_1^{-1} [I(x_M)/ x_M] \nonumber \\
&=& 
1.783\times 10^{-7} 
C_1^{-5} C_2^{-16/7} D_8^{-8/7}
\left({x_{M}\over7.314}\right)^{-2}
\left[{x_{M} I(x_{M})\over0.7811}\right]\,,
\end{eqnarray}
so the source is optically thin above $x_M = 1.766 \times 10^{-3}$, i.e., for
$\nu$ exceeding $34.52$ GHz.  Then 
the photon energy density 
\begin{eqnarray}
U&\simeq& {{\cal L}_{\rm syn}/ 4 \pi c\ R^2} = 127.9 C_1^{-4} C_2^{-12/7} 
D_8^{-6/7} \times\nonumber \\
&&\left({x_{M}\over7.314}\right)^{-5/7}
\left({x_{M} I(x_{M})\over0.7811}\right)^{3/7}
\left({\nu_{14}\over 1.429}\right)^{5/7} 
\left({F_\nu\over 7\ {\rm mJy}}\right)^{-3/7} 
{\rm ergs\ cm}^{-3}
\end{eqnarray}
is more than 3 times lower than that of the magnetic field, justifying our neglect of the self-Comptonization 
cooling\footnote{The self-Comptonization effects may be incorporated into the model by replacing the $B^2$
with $B^2+8\pi U$ in the electron cooling and source luminosity terms.}.

The self-Comptonization flux density is then given by (Blumenthal \& Gould 1970)
\begin{eqnarray}
F_X(\nu) \simeq (2\pi e^7 n^2 B R^4/ 3 \sqrt{3} m_e^3 c^6 D^2) 
G({\nu/4\nu_c\gamma_c^2}) 
= 2.121\ n_7^2\ B_1\ R_{12}^4\ D_8^{-2} 
G({\nu/4\nu_c\gamma_c^2}) 
\mu{\rm Jy}\,,
\end{eqnarray}
where 
\begin{equation}
G(z) = z\int_{0}^\infty {\rm d} x\int_0^1{\rm d}y\;
\exp(-x) I({zx^{-2}y^{-1}})(2\ln{y}+1-2y+
y^{-1})\,,
\end{equation}
the low limit of the integral over $x=\gamma/\gamma_c$ has been extended from $1/\gamma_c$ to 0, which 
simplifies the formula and introduces a less than $0.2\%$ error, and $z=\nu/4\nu_c\gamma_c^2$. 
Similarly one can define the spectral index
\begin{equation}
\alpha_X \equiv {{\rm d}\ln(\nu F_X)/ {\rm d}\ln\nu} = 1+ {{\rm d}\ln G(z)/{\rm d}\ln z} \,.
\label{alphax0}
\end{equation}
The normalized power spectrum $zG(z)$ and $\alpha_X$ are depicted by thin lines in Figure \ref{fig1.ps} 
(left).

For the above mentioned NIR flare, the model predicts that in the {\it Chandra} and {\it XMM} bands ($\nu^X = 
\nu^X_{18}10^{18}$ Hz, which corresponds to $\sim 4.2$ keV) 
\begin{eqnarray}
z &=& 0.28\
C_1^{-2} C_2^{-8/7} D_8^{-4/7}\ \nu^X_{18}
\left({x_{M}\over7.314}\right)^{13/7}
\left({x_{M} I(x_{M})\over0.7811}\right)^{2/7}
\left({\nu_{14}\over 1.429}\right)^{-13/7} 
\left({F_\nu\over 7\ {\rm mJy}}\right)^{-2/7}\,, \label{alphax}\\
\alpha_X(z) &=& 1.1\,, \\  
F_X(z)&=& 0.0033\
C_1^{-2} C_2^{-4/7} D_8^{-2/7}
\left({G(z)\over 0.58}\right)\times \nonumber \\
&& 
\left({x_{M}\over7.314}\right)^{3/7}
\left({x_{M} I(x_{M})\over0.7811}\right)^{-6/7}
\left({\nu_{14}\over 1.429}\right)^{-3/7} 
\left({F_\nu\over 7\ {\rm mJy}}\right)^{6/7} \mu{\rm Jy}\label{fx}
\,,
\end{eqnarray}
which is about 5 times below the quiescent flux level. We therefore do not expect a detectable X-ray flux 
accompanying this flare. From equations (\ref{alphax}) and (\ref{fx}), one obtains the correlation between 
the NIR and X-ray spectroscopic observations. For a given X-ray spectral index, which is a function of $z$, 
Equation (\ref{alphax}) gives the dependence of the NIR flux density on $x_{M}$, which determines $\alpha$. 
For a given X-ray flux density $F_X$, one obtains $x_{M}$ as a function of  $z$ by eliminating the 
$F_\nu$ terms in the two equations. This gives $F_\nu$ as a function of $\alpha$ with $x_{M}$ as a parametric 
variable. These two relations are plotted in Figure \ref{fig1.ps} (right) for $C_1=C_2=D_8=1.0$.
Here $F_X$ is in units of Sagittarius A*'s quiescent flux density of $0.015\ \mu$Jy at $4.2$ keV. 

\section{Discussion and Conclusions}
\label{dis}

First, we notice that for weak NIR flares with very hard spectra (to the bottom-right of the figure), 
there is little X-ray emission and that X-ray flares are expected to accompany NIR flares with soft spectra (to 
the left). These predictions may be readily tested via a statistical study of all the observed flare events
(Hornstein et al. 2002) and explain why the occurrence rate of NIR 
flares is more than two times greater than that of the X-ray events. For NIR flares with very hard spectra, 
$\alpha_X$ is also large (corresponding to a hard spectrum), 
we expect hard X-ray/$\gamma$-ray emission, especially during bright NIR flares (to the 
top-right). 

Second, brighter X-ray flares (to the top-left) tend to have higher NIR fluxes and softer NIR 
spectra. $\alpha_X$ is also small to the left.
We expect little $\gamma$-ray emission accompanying weak NIR flares with 
very soft spectra (to the bottom-left). 

Third, for a given $\alpha$, $F_X$ is roughly proportional to $F_\nu$. The spectral index $\alpha_X$ 
correlates with $F_{\nu}$ for low values of $\alpha$, but this correlation almost vanishes for positive 
$\alpha$'s. For low values of $\alpha$, we also expect a correlation between $F_X$ and $\alpha_X$. 

Finally, the results are very sensitive to the values of $C_1$ and $C_2$. The dashed lines in the figure 
indicates X-ray flares with $C_1=0.5$. These are quite different from the lines with 
$C_1=1.0$. For the referenced NIR flare, the X-ray flux will be detectable with {\it Chandra} for $C_1=0.5$. 
Thus $C_1$ and $C_2$ may be determined accurately with SSO.
 
Some of these predictions are shown explicitly in the left panel of Figure \ref{fig2.ps}, where the solid 
line is the model spectra of the synchrotron and self-Comptonization components for the NIR flare studied 
above. Although there is little X-ray emission, we expect $\gamma$-ray emission up to a few MeV during the 
flare. The model is also applied to the brightest flare observed simultaneously in the NIR and X-ray bands by 
Eckart et al. during their multi-wavelength campaign in 2004. The NIR and X-ray fluxes correlated well during 
the flare and the peak flux densities $F_\nu = 7$mJy and $F_X=0.223\ \mu$Jy. The spectral indexes of the flare 
have not been published yet. Because the X-ray flux is high, we expect soft NIR spectra. The dotted (with 
$C_1=C_2=1.0$) and dashed (with $C_1=2.2$, $C_2=0.74$) lines are for this flare. The right panel shows 
how the NIR and X-ray spectral indexes depend on $C_1$ and $C_2$. The cross and square correspond to the 
dotted and dashed lines in the left panel, respectively. These show that for given 
flux densities the spectral indexes are very sensitive to the values of $C_1$ and $C_2$. 
Because both $C_1$ and $C_2$ are dimensionless 
quantities, we expect them to be on the order of 1. The model therefore predicts that the NIR emission has a 
soft spectrum with $\alpha<0$. 

\begin{figure}[bht] 
\begin{center}
\includegraphics[height=8.4cm]{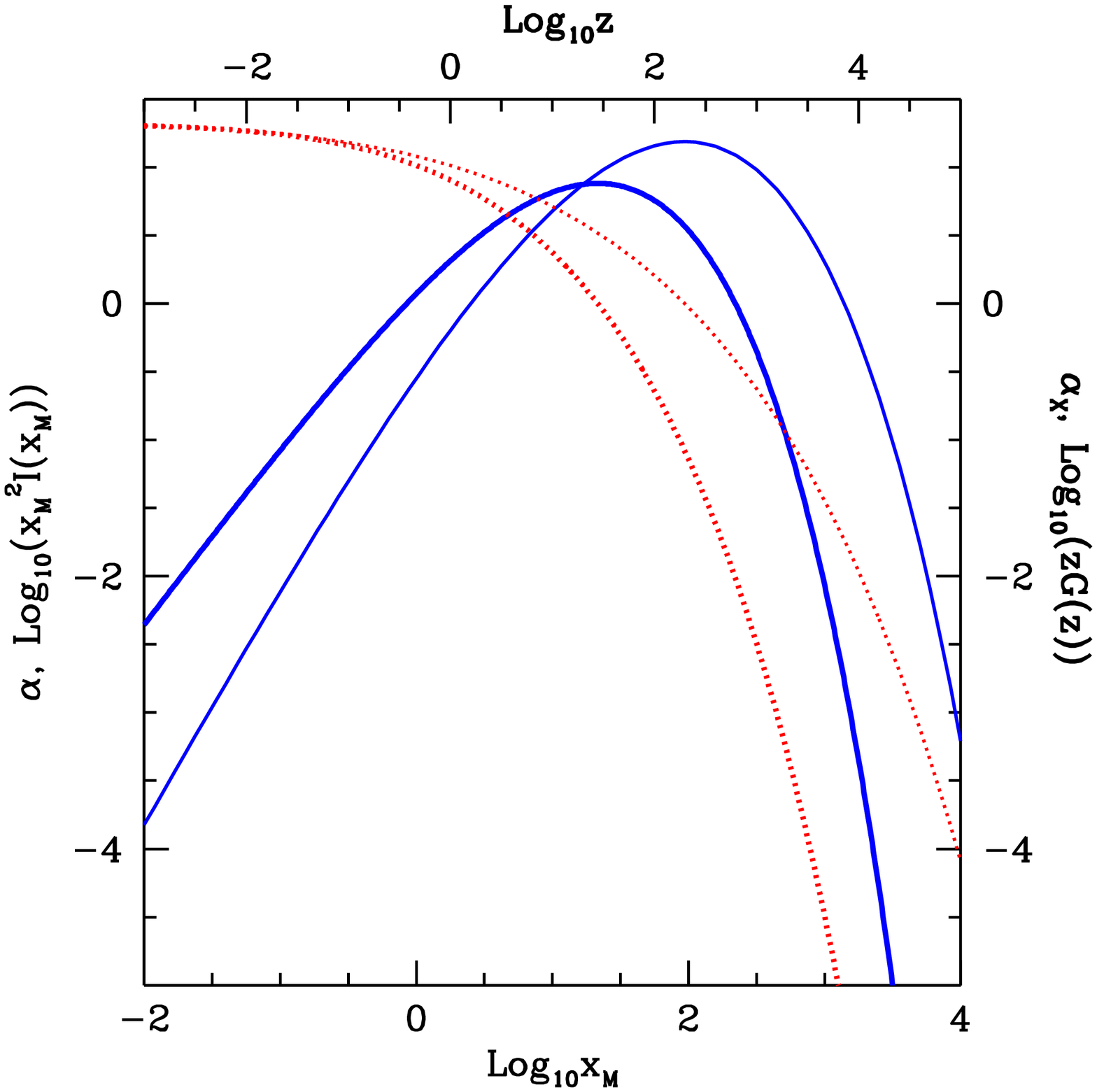}
\hspace{-0.6cm}
\includegraphics[height=8.4cm]{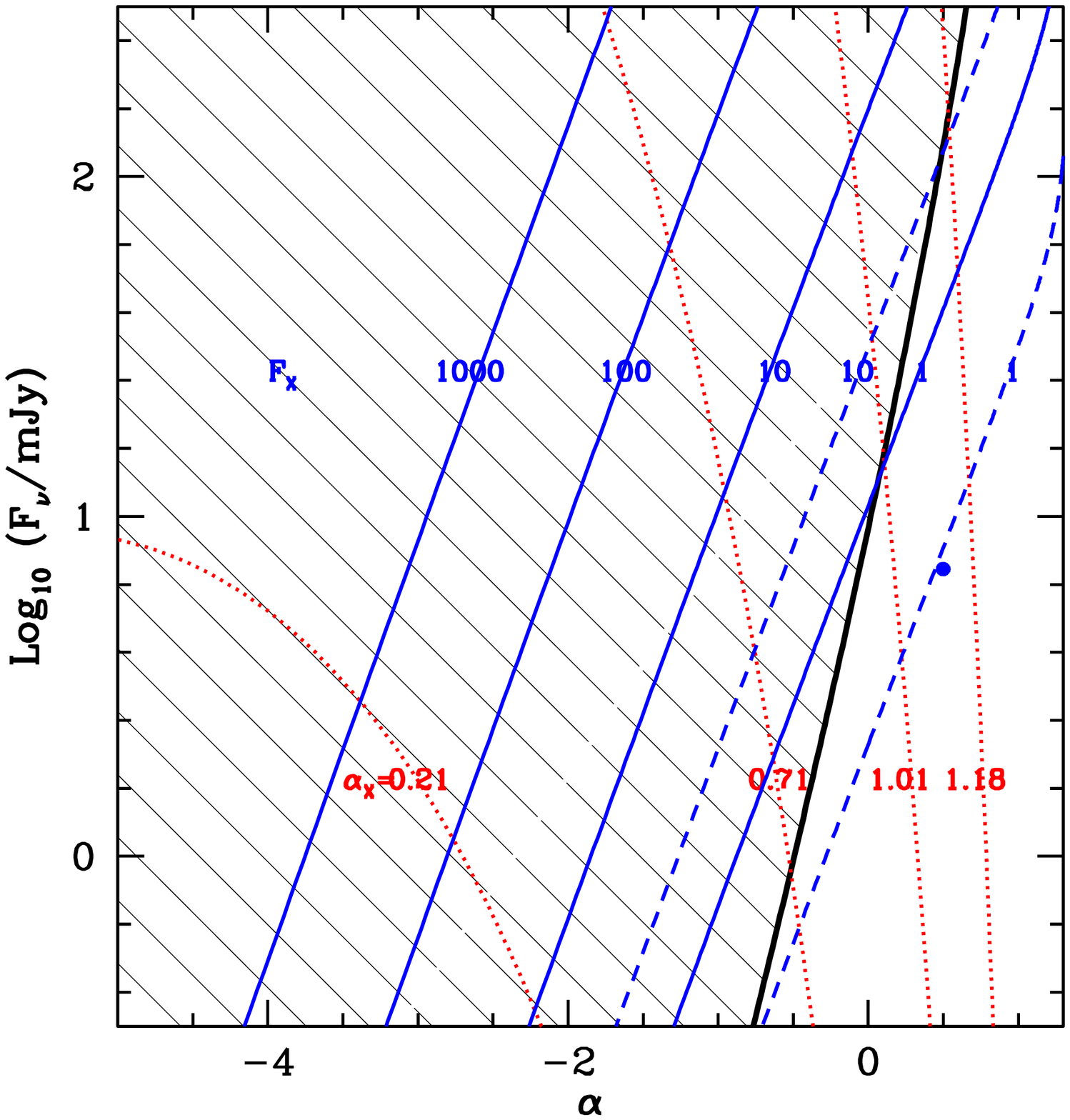}
\end{center}
\caption{
{\it Left:} Thermal synchrotron (thick) and self-Comptonization (thin) power spectral index (dotted) and 
normalized energy flux density $\nu F_\nu$ (solid) as functions of their normalized frequencies. 
{\it Right:} Correlation between the NIR ($\nu_{14}=1.429$) and X-ray ($\nu_{18}=1.0$) spectral indexes and 
flux densities for $C_1=C_2=D_8=1.0$. The horizontal and vertical axises are for the NIR spectral index and 
flux density, respectively. The dotted and solid lines indicate constant X-ray spectral index and flux 
density (in units of the quiescent level of $0.015\mu$Jy) with the corresponding value labeled on the lines. 
The dashed lines are similar to the solid lines but with $C_1=0.5$. The dot indicates the model studied in 
the paper. The steady-state solution may not be valid in the shaded region. See text for details.
}
\label{fig1.ps}
\end{figure}

\begin{figure}[bht] 
\begin{center}
\includegraphics[height=8.4cm]{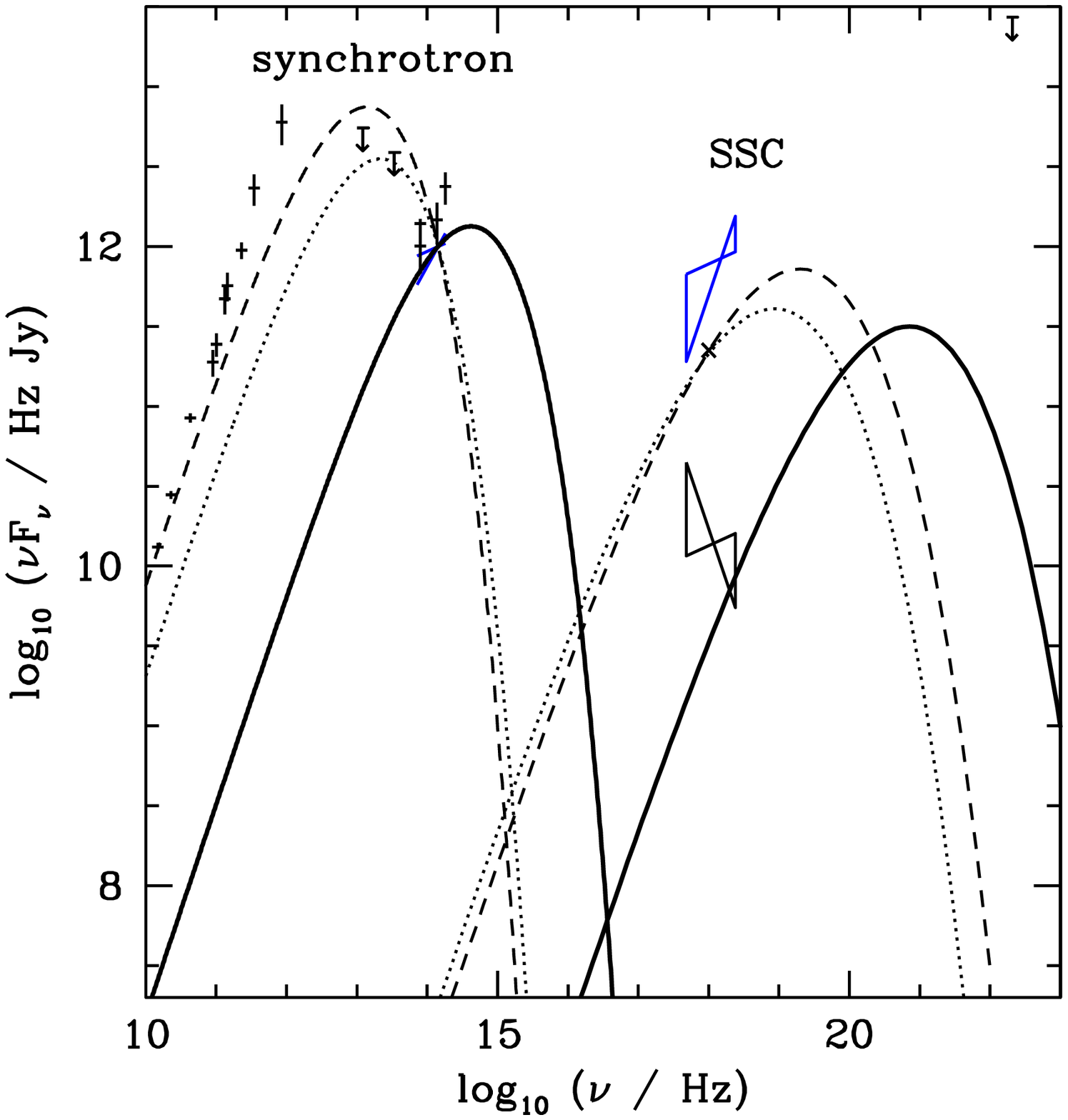}
\hspace{-0.6cm}
\includegraphics[height=8.4cm]{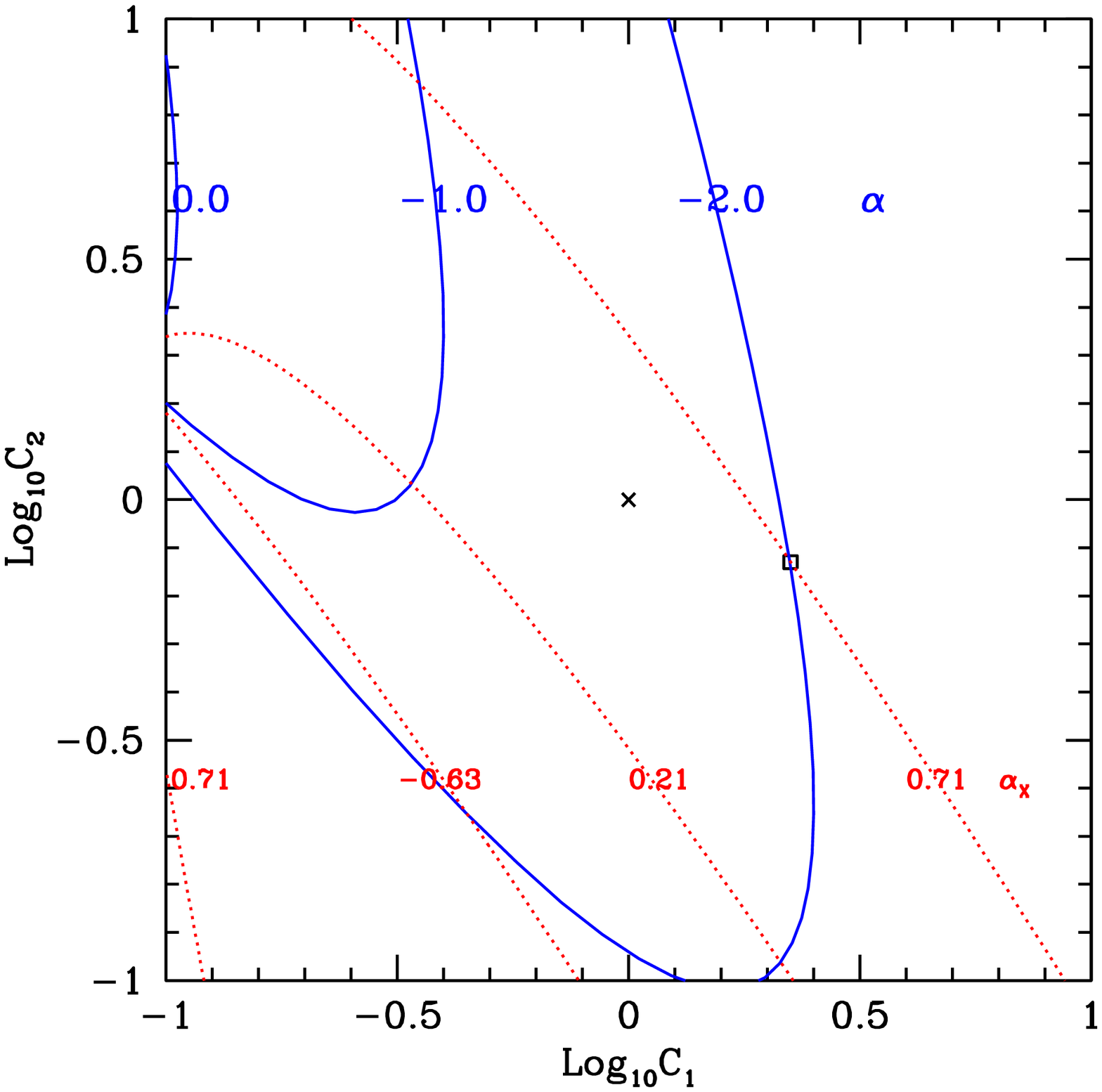}
\end{center}
\caption{
{\it Left:} Model fit to the NIR flare studied in the paper (solid line) and the peak flux densities of 
the brightest flare observed simultaneously in the NIR and X-ray bands by Eckart et al. (2006). For the 
latter, $F_\nu = 7$ mJy and $F_X = 0.223\ \mu$Jy. $C_1 = C_2 = 1.0$ for the dotted line, and $C_1=2.2$, 
$C_2=0.74$ for the dashed line. Note that the effect of self-absorption is not included here. The {\it 
Right} panel shows the dependence of the model predicted NIR ($\alpha$) and X-ray ($\alpha_X$) spectral 
indexes on $C_1$ and $C_2$ for this flare. We expect a soft NIR spectrum with $\alpha<0.0$ for reasonable 
values of $C_1$ and $C_2$.
}
\label{fig2.ps}
\end{figure}

For the referenced NIR flare the electron acceleration time
\begin{equation}
\tau_{\rm ac} = 5.845\,C_1^3C_2^{12/7}D_8^{6/7}
\left({x_{M}\over7.314}\right)^{5/7}
\left({x_{M} I(x_{M})\over0.7811}\right)^{-3/7}
\left({\nu_{14}\over1.429}\right)^{-5/7} 
\left({F_\nu\over 7\ {\rm mJy}}\right)^{3/7}\
{\rm mins}
\end{equation}
is shorter than the dynamical time of $\tau_{\rm dy}\sim 20$ mins, which is consistent with the observed 
flare variation time, justifying our usage of the steady-state solution, 
which, however, may not be valid when $\tau_{\rm ac}>\tau_{\rm dy}$, as indicated by the shaded area in Figure 
\ref{fig1.ps} (right). Note that for the NIR flares with softer spectra we studied before (Liu et al. 2006), 
$\alpha<-2.0$, the steady-state solution therefore may not be applicable. In the most general cases, both 
$\tau_{\rm ac}$ and $\tau_0$ are functions of $t$, which are determined by the large scale MHD processes.
Time dependent solutions of equation (\ref{kinetic}) $N(\gamma, t)$ are needed. However, we don't expect the 
electron distribution to be radically different from Maxwellian. By introducing a time-dependent 
$\gamma_c(t)$, the results here may be readily generalized to incorporate the 
time evolution of the system and applied to the accretion flow in the quiescent state when advection can be 
important. One may also have to take into account the Doppler and gravitational 
effects as the emitting plasma is likely moving on a Keplerian orbit near the black hole (Bromley et al. 
2001; Broderick \& Loeb 2005). These are beyond the scope of the current investigation.

Since the discovery of NIR and X-ray flares in Sagittarius A*, several models have been proposed (Markoff 
et al. 2001; Liu \& Melia 2002). Given the uncertainties in the electron acceleration mechanism, earlier 
models prescribe either the electron distribution (Yuan et al. 2004) or the turbulence spectrum (Liu et al. 
2004) with several free parameters resulting in all kinds of emission spectra. In a previous work (Liu et al. 
2006), we showed that Sagittarius A*'s flare activity might be explained with thermal synchrotron and 
self-Comptonization emission and all 
the model parameters, namely $B$, $n$, $R$, and $\gamma_c$, could be readily determined by SSO. In this paper 
we have shown that the electrons are likely heated by turbulence, which produces a Maxwellian distribution and 
predicts two time-insensitive constraints on the plasma properties making the model 
testable with SSO. This is the first model that can be tested with current instrument capability. 
We have provided general expressions for the emission spectra and demonstrated how properties of 
the flaring plasma 
may be inferred from the spectroscopic observations or simultaneous flux measurements in both bands. 
The model may therefore provide powerful tools for studying physical processes near the event horizon of black 
holes. A time-dependent treatment of this problem will make it possible to compare MHD simulations of the 
black hole accretion with the observed flare activity in Sagittarius A*, which will eventually lead to a 
measurement of the black hole's spin after the instability is identified and light propagation effects are 
incorporated properly. (Note that for the referenced NIR flare the electron Lorentz factor $\gamma_c\sim 200$ 
already suggests that the gravitational energy release beyond the last stable orbit of a Schwarzschild black 
hole is not sufficient to power the flare.) The formulae for the thermal synchrotron and self-Comptonization 
emission are valid so long as optically thin synchrotron emission dominates the radiative output and may 
therefore be applicable to other astrophysical sources, particularly Low-Luminosity-AGNs.

\acknowledgments

We thank Mark Morris, Seth Hornstein, and Marco Fatuzzo for useful discussion. 
This work was funded in part under the auspices of the U.S.\ Dept.\ of Energy, and supported 
by its contract W-7405-ENG-36 to Los Alamos National Laboratory.
This research 
was partially supported by NSF grant ATM-0312344, NASA grants NAG5-12111, NAG5 11918-1 (at Stanford), NSF 
grant AST-0402502 (at Arizona), and NSF grant PHY99-07949 (at KITP at UCSB). 
FM is very grateful to the University of Melbourne for its support (through a Miegunyah Fellowship).

\newpage
\begin{thebibliography}{}

\bibitem[Baganoff et al. 2001] {Baganoff01} Baganoff, F. K., et al. 2001, Nature,
413, 45
\bibitem[Baganoff et al. 2003]{Baganoff03} Baganoff, F. K., et al. 2003, ApJ, 591, 891
\bibitem[Belanger et al. 2005]{Belanger06} Belanger, G., et al. 2005, ApJ, 635, 1095
\bibitem[Belanger et al. 2006]{Belanger05} Belanger, G., et al. 2006, \apj, submitted
\bibitem[Blandford \& Eichler 1987]{Bland87} Blandford, R., \& Eichler, D. 1987, Phys. Report, 154, 1
\bibitem[Blumenthal \& Gould 1970]{Blum70} Blumenthal, G. R., \& Gould, R. J. 1970, Rev. of Mod. Phys. 42, 237
\bibitem[Broderick \& Loeb 2005]{Brod05} Broderick, A. E., \& Loeb, A. 2005, MNRAS, 363, 353 
\bibitem[Bromley et al. 2001]{brom01} Bromley, B., Melia, F., \& Liu, S. 2001, ApJ, 555, L83
\bibitem[Dermer, Miller, \& Li 1996]{dermer96}Dermer, C. D., Miller, J. A., \& Li, H. 1996, ApJ, 456, 106
\bibitem[Eckart et al. 2004] {Eckart04} Eckart, A., et al. 2004, A\&A, 427, 1
\bibitem[Eckart et al. 2006] {Eckart06} Eckart, A., et al. 2006, A\&A, in press
\bibitem[Eisenhauer et al. 2005] {Eisen05} Eisenhauer, F., et al. 2005, \apj, 628, 246
\bibitem[Fermi 1949]{Fermi49} Fermi, E. 1949, Phys. Rev. 75, 1169
\bibitem[Genzel, et al. 2003]{Genzel03} Genzel, R. et al. 2003, Nature, 425, 934
\bibitem[Ghez et al. 2004]{Ghez04} Ghez, A. M., et al. 2004, ApJ, 601, L159
\bibitem[Ghez et al. 2005]{Ghez05} Ghez, A. M., et al. 2005, ApJ, 635, 1087 
\bibitem[Gillessen et al. 2006]{Gil06} Gillessen S., et al. 2006, ApJ, 640, 163L
\bibitem[Goldwurm et al. 2003]{Gold03} Goldwurm, A. et al. 2003, \apj, 584, 751
\bibitem[Hornstein et al. 2006]{Horn06} Hornstein, S., et al. 2006, ApJ, in preparation 
\bibitem[Hornstein et al. 2002]{Horn02} Hornstein, S., et al. 2002, ApJ, 577, 9 
\bibitem[Liu \& Melia (2002)] {Liu02} Liu, S., \& Melia, F. 2002, \apjl, 566, L77
\bibitem[Liu, Melia, \& Petrosian 2006]{Liu06} Liu, S., Melia, F., \& Petrosian,
V. 2006, \apj, 636, 798
\bibitem[Liu, Petrosian, \& Melia 2004] {Liu04} Liu, S., Petrosian, V., \& Melia, F.
2004, \apjl, 611, L101
\bibitem[Mahadevan et al. 1996]{Maha96} Mahadevan, R., Narayan, R., \& Yi, I. 1996, \apj, 465, 327
\bibitem[Markoff et al. 2001]{Markoff01} Markoff, S., Falcke, H., Yuan, F., \& Biermann, P. L. 2001, 379, L13
\bibitem[Melia 2006]{Melia06} Melia, F. 2006, {\it The Galactic Supermassive Black Hole},
Princeton University Press.
\bibitem[Melia et al. (2000)]{Melia00} Melia, F., Liu, S., \& Coker, R. 2000, \apjl, 545, L117
\bibitem[Melia et al. (2001)]{Melia01} Melia, F., Liu, S., \& Coker, R. 2001, \apj, 553, 146
\bibitem[Nayakshin, Cuadra, \& Sunyaev 2004]{Nayak04} Nayakshin, S., Cuadra, J., \& 
Sunyaev, R. 2004, A\&A, 413, 173
\bibitem[Park \& Petrosian 1995]{Park95} Park, B. T., \& Petrosian, V. 1995, ApJ, 446, 699
\bibitem[Petrosian 1981]{Petr81} Petrosian, V. 1981, \apj, 251, 727
\bibitem[Petrosian \& Liu 2004]{Petr04} Petrosian, V., \& Liu, S. 2004, ApJ, 610, 550
\bibitem[Porquet et al. 2003]{Porquet03} Porquet, D., et al. 2003, A\&A, 407, L17 
\bibitem[Schlickeiser 1984]{Sch84} Schlickeiser, R. 1984, A\&A, 136, 227
\bibitem[Tagger \& Melia 2006]{Tag06} Tagger, M., \& Melia, F. 2006, ApJ, 636, L33
\bibitem[Yuan et al. 2003]{Yuan03} Yuan, F., Quataert, E., \& Narayan, R., 2003, 598, 301
\bibitem[Yuan et al. 2004]{Yuan04} Yuan, F., Quataert, E., \& Narayan, R., 2004, 606, 894
\bibitem[Yusef-Zadeh et al. 2006]{Yu06} Yusef-Zadeh, F., et al. 2006, ApJ, in press.
\bibitem[Zhao et al. 2004]{Zhao04} Zhao, J. H., Herrnstein, R. M., Bower, G. C., Goss, W. M., 
\& Liu, S. M. 2004, ApJ, 603, L85

\end{thebibliography}
\end{document}